# The Effect of Intrinsic Crumpling on the Mechanics of Free-Standing Graphene

Ryan J.T. Nicholl[1], Hiram J. Conley[1], Nickolay V. Lavrik[2], Ivan Vlassiouk[3], Yevgeniy S. Puzyrev[1], Vijayashree Parsi Sreenivas[1], Sokrates T. Pantelides[1], and Kirill I. Bolotin[1,4*]

[1]Department of Physics and Astronomy, Vanderbilt University, Nashville, TN 37235, USA
[2]Center for Nanophase Materials Sciences, Oak Ridge National Laboratory, Oak Ridge, TN 37831, USA
[3]Energy & Transportation Science Division, Oak Ridge National Laboratory, Oak Ridge, TN 37831, USA
[4]Department of Physics, Freie Universität Berlin, Arnimallee 14, 14195 Berlin, Germany

*kirill.bolotin@vanderbilt.edu

**Free-standing graphene is inherently crumpled in the out-of-plane direction due to dynamic flexural phonons and static wrinkling. We explore the consequences of this crumpling on the effective mechanical constants of graphene. We develop a sensitive experimental approach to probe stretching of graphene membranes under low applied stress at cryogenic to room temperatures. We find that the in-plane stiffness of graphene is 20 – 100 N/m at room temperature, much smaller than 340 N/m (the value expected for flat graphene). Moreover, while the in-plane stiffness only increases moderately when the devices are cooled down to 10 K, it approaches 300 N/m when the aspect ratio of graphene membranes is increased. These results indicate that softening of graphene at temperatures <400 K is caused by static wrinkling, with only a small contribution due to flexural phonons. Together, these results explain the large variation in reported mechanical constants of graphene devices and paves the way towards controlling their mechanical properties.**

What is the mechanical stiffness of free-standing monolayer graphene? The answer to this question may appear trivial: graphene is universally modeled as a flat sheet with in-plane stiffness (the ratio of in-plane stress and strain) $E_{2D} = 340$ N/m and vanishing bending rigidity. This record stiffness caused by strong carbon-carbon bonds[1,2] is confirmed by multiple experiments[3,4] and is consistent with the bulk Young's modulus of graphite $E_{3D} \sim 1$ TPa $\sim E_{2D}/t$, where $t \sim 0.335$ nm is graphite interlayer spacing.[5] However, a number of experiments[4,6,7] find largely varying values of in-plane stiffness. One possible reason for this is that under realistic experimental conditions, free-standing (or even substrate-supported) graphene is never flat but is inevitably crumpled in the out-of-plane direction. This crumpling predominately originates from two different mechanisms. The first mechanism is static wrinkling, likely due to uneven stress at the boundary of graphene produced during device fabrication. Static wrinkles are quasi-periodic undulations with heights up to 30 nm.[8] The second mechanism is dynamic crumpling due to out-of-plane flexural phonons. Such flexural phonons are responsible for the negative thermal expansion coefficient of graphene[9] and affect its electrical and thermal conductivity.[10,11] It stands to reason that the crumpling of graphene should strongly perturb its effective mechanical properties and in particular decrease its in-plane stiffness.[12,13]

At the same time, the interplay between crumpling and mechanical properties of graphene remains virtually unstudied experimentally, despite its obvious significance for fundamental understanding of mechanics of 2D materials as well as applications in graphene-based NEMS devices. We believe that the likely culprit is the technique used in the majority of the experiments probing graphene mechanics – atomic force microscopy (AFM) nanoindentation.[3,7,14,15] In that approach, effective mechanical constants close to that of flat graphene can be found since very large non-uniform stress applied to graphene can "flatten out" any crumpling. Additionally, small sample sizes ($\sim 1$ μm) used in many experiments can further suppress crumpling. Recently, Ruiz-Vargas et al.[16] reported decreased in-plane stiffness associated with crumpling of graphene. However, the mechanism of the crumpling, the factors affecting it, and the contribution to crumpling due to defects and grain boundaries in CVD graphene remained unknown.

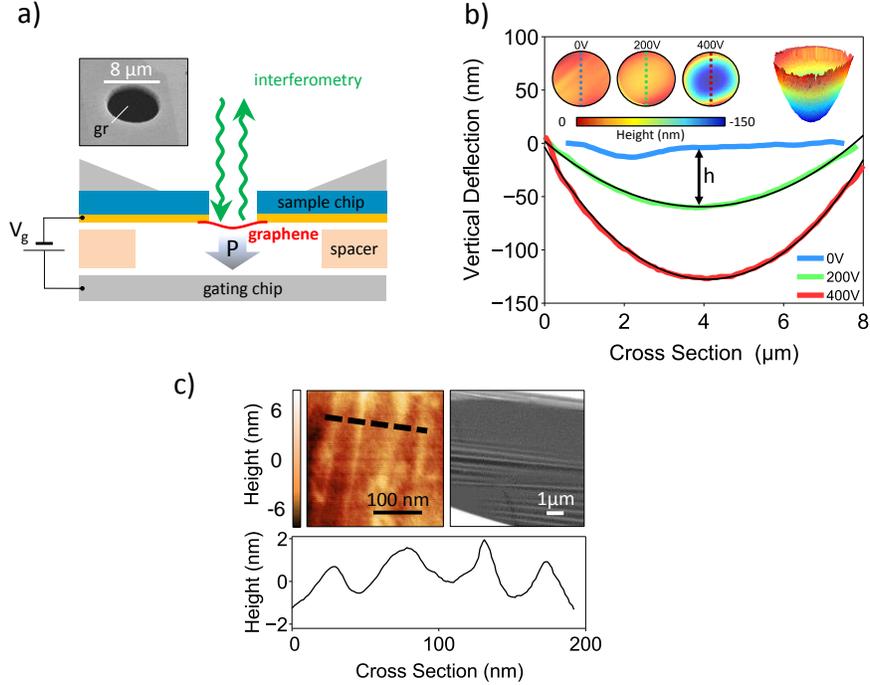

**Figure 1: Experimental setup.** *(a)* Device schematic. Inset: SEM image of a representative free-standing graphene membrane (the scale bar is 8 μm). *(b)* Cross-sections of a graphene membrane at various applied voltages. Height data obtained from interferometric profilometry corresponding to these cross-sections are shown in the inset. Also shown is a 3D view of the data at $V_g = 400$ V. *(c)* AFM measurements of graphene membrane showing nanometer-scale static wrinkles (left, the scale bar is 100 nm). A cross section of the AFM data is shown in the bottom panel. Wrinkling is also evident on the high-angle tilted SEM image (right, the scale bar is 1 μm).

Here, we quantitatively investigate the effects of crumpling on the effective mechanical stiffness of graphene. To accomplish this, we develop a non-contact approach based on interferometric profilometry. This allows us to study the mechanics of graphene at cryogenic to room temperatures in a controlled geometry with uniform loading. We confirm that out-of-plane crumpling softens $E_{2D}$ to values of as low as ~20 N/m. Furthermore, by performing temperature-dependent measurements and studying devices of various geometries, we separately probe the contributions of flexural phonons and static wrinkles. Static wrinkling is found to be the dominant mechanism determining the effective mechanical constants in our devices. Finally, we discover an approach to suppress static wrinkling by in-situ changing the membrane's geometry.

## Results
### Experimental setup

Our experimental setup is shown in Fig. 1a. At its heart is a suspended graphene membrane that is actuated by applying an electrical bias between it and a silicon "gating chip" underneath and whose deflection is monitored via interferometric profilometry. Graphene membranes suspended over holes in silicon nitride ($Si_3N_4$) with diameter ranging between 7.5 μm and 30 μm form the "sample chip". Pristine residue-free graphene is seen in the high resolution SEM imaging of the sample chip (Fig. 1a, inset). The sample chip is placed onto the gating chip consisting of degenerately doped silicon coated with 2 μm of $SiO_2$. To provide additional electrical insulation, a 7.5 μm thick Kapton film with a hole punched in the center is inserted between sample and gating chips. The entire structure is then mechanically clamped, resulting in an average separation of 15 μm between the graphene and the gating chip. Finally, separate electrical contacts are made to the graphene and the gating chip. The entire device structure is placed inside an optical cryostat in vacuum better than $10^{-5}$ Torr at temperatures between 4 K and 400 K.

Graphene was electrostatically pressured by applying a voltage $V_g$ between the graphene and the gating chip. The pressure applied to graphene can be evaluated as $P = \varepsilon_0 V_g^2 / 2d^2$; where $\varepsilon_0$ is the vacuum permittivity, and $d$ is the separation between graphene and gate as determined by interferometric profilometry (discussed below). The applicability of parallel-plate capacitor approximation is justified since $d \sim 10 - 20$ μm is much larger compared to the maximum deflection of graphene ($\sim 600$ nm). The maximum $V_g$ that can be applied without dielectric breakdown is $\sim 2000$ V, which allowed us to reach maximum pressures around 30 kPa; the uncertainty in $P$ is below 5% for all voltages.

The deflection of the graphene membranes under applied pressure was determined via interferometric profilometery. In this technique, a profile of the surface $h(x, y)$ is determined with sub-nanometer precision in the out-of-plane direction and sub-micron resolution in the in-plane direction by detecting the phase shift of light reflected from the sample surface. Large graphene-gate separation $d$ helps to maximize the interferometric signal from nearly transparent graphene. When $d$ is large, only the signal reflected from graphene is detected since the gating chip is out of focus. A separate measurement is performed to find $d$ by deliberately sweeping the focus from the sample to gating chip.

Our method of probing the mechanical properties of graphene has several critical advantages over AFM-based nanoindentation and other techniques. First, graphene deflection is measured via a non-contact approach. This means that the membrane morphology is not disturbed with a sharp tip that applies non-uniform stress. Second, the height data from the entire membrane is recorded at the same time. This means we can find the true maximum center point deflection, $h$, and verify that the membrane is deflecting symmetrically. Third, the pressure is applied uniformly, allowing us to use simple and reliable models to extract mechanical constants.

Fourth, the optical nature of the technique allows simple characterization of devices inside an optical cryostat at low temperatures. At the same time, we note that our method is not a replacement for AFM nanoindentation. Our approach operates in the regime of low applied pressure and cannot be used to determine breaking strength of 2D materials.

**Determination of mechanical constants of graphene membranes**

The in-plane stiffness of graphene at room temperature is extracted from measured membrane profiles $h(x, y)$ vs. known applied pressure $P$ (Fig. 1b). Through simple geometric considerations, we determined the radial in-plane stress $\sigma_r$ and radial strain $\varepsilon_r$:[17]

$$\sigma_r = \frac{Pa^2}{4h} \qquad \varepsilon_r = \frac{2h^2}{3a^2} \qquad (1)$$

Here $a$ is the radius of a graphene membrane, and $h$ is its center-point deflection determined by fitting the experimental data to a sphere. In the majority of measured devices we observe a linear relationship between $\varepsilon_r$ and $\sigma_r$ (Fig. 2a), allowing us to determine the in-plane stiffness of graphene. It is given by $E_{2D} = (1 - \nu)\sigma_r / \varepsilon_r$, where $\nu \sim 0.165$ is the Poisson ratio for graphene.[18] While in realistic devices both strain and stress vary slightly throughout the device, our finite element modeling (FEM) confirms that the exact numerical solution for $E_{2D}$ does not deviate more than 10% from the simple analytical estimate above (Supplementary Fig. 1). This is within the uncertainty of extracting $E_{2D}$ from our data. We also note that the obtained $E_{2D}$ agrees with the value obtained by fitting $P(h)$ data to the bulge-test equation $P = 4\sigma_0 h / a^2 + 8 E_{2D} h^3 / 3(1 - \nu)a^4$ (Fig. 2a, Inset) commonly used in thin film mechanics measurements.[17,19] In all 26 measured monolayer CVD graphene membranes (Fig. 2b) we find $E_{2D} = 35 \pm 29$ N/m, consistent with previous work.[16] In a few devices we observed pronounced non-linear dependence of $\sigma_r(\varepsilon)$, with $E_{2D} = (1 - \nu) d\sigma_r / d\varepsilon_r$ increasing from less than 10 N/m at low stress to 50 N/m at higher stress (Fig. 2c).

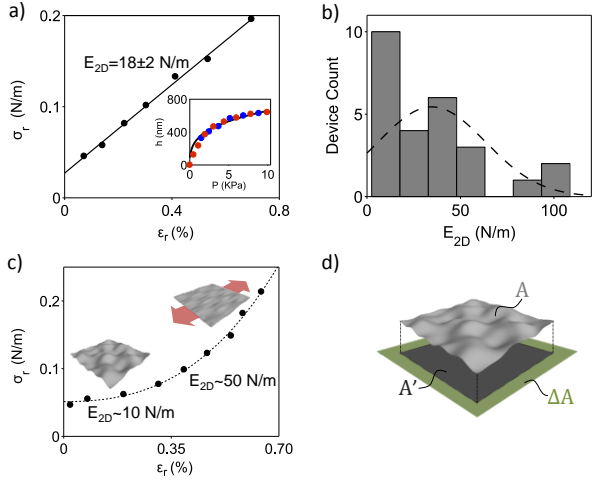

**Figure 2: Mechanics of graphene membranes at room temperature.** *(a)* Stress ($\sigma_r$) *vs.* strain ($\varepsilon_r$) dependence for a typical device. The in-plane stiffness $E_{2D}$ is extracted from the slope of linear fit to these data (black line). Inset: Raw center-point deflection, $h$, *vs.* pressure, $P$, data used for calculation of stress and strain (red: loading cycle, blue: unloading cycle) along with a fit to the bulge-test equation (black line). *(b)* Histogram of $E_{2D}$ for all measured CVD graphene devices. *(c)* A non-linear stress-strain curve seen in a minority of devices. The dashed line is a guide to the eye. *(d)* A cartoon view of a crumpled membrane.

We performed numerous consistency checks to rule out possible measurement artifacts. First, we observed no hysteresis in $P(h)$ data between loading and unloading cycles (Fig. 2a, Inset). This establishes that graphene is not slipping against the substrate. Second, we observed similarly soft $E_{2D}$ for CVD graphene (Fig. 2b; grain size ~50 μm, bigger than the membrane size) and exfoliated graphene ($E_{2D}$~50 − 80 N/m in two devices). This confirms that $E_{2D}$ in our experiments is not affected by the grain boundaries in graphene, which is consistent with conclusions from previous experiments.[15] Third, we cross-checked our results against the measurements obtained via AFM nanoindentation (Supplementary Fig. 3). In the regime of low loading forces < 300 nN, nanoindentation measurements on the same devices yielded $E_{2D}$ consistent with optical profilometry measurements. It is important to note that AFM nanoindentation pushes graphene towards the substrate, while electrostatic loading pulls graphene away from it. Similarity in $E_{2D}$ values obtained for opposite loading directions confirms that interaction of graphene with the sidewalls does not affect the measured $E_{2D}$. Finally, simple estimates show that the organic residues that may remain on graphene after the fabrication process are unlikely to affect $E_{2D}$. A uniform residue layer with Young's modulus of ~2 GPa [20] and thickness < 5 nm is expected to be at least 100 times softer compared to graphene.

**Probing the effects of static and dynamic crumpling**

In the remaining part of the manuscript, we demonstrate that the observed decrease in stiffness is due to the crumpled nature of graphene. Indeed, static wrinkles with wavelength ~50 nm and average amplitude ~1 nm are observed in our samples via AFM (Fig. 1c, Left/Bottom). Somewhat larger micron-scale features are seen in a minority of membranes, as shown in the Scanning Electron Microscopy (SEM) images (Fig. 1c, Right). Flexural phonons are invariably present in graphene at room temperature. Signatures of such flexural phonons have been observed in transmission electron microscopy imaging of graphene.[21] Similarly, uncontrolled stress and hence static wrinkling is always present in experimentally available free-standing graphene specimens.[22]

To understand intuitively how crumpling due to flexural phonons or wrinkles can affect mechanics of graphene, one only has to consider a simple analogy – a sheet of paper. When flat, such a sheet is very stiff. However, once the same sheet is crumpled, it becomes very easy to stretch. The reason for this behavior is that stretching of a crumpled sheet mostly flattens and bends it. In contrast, stretching of a flat sheet strains it locally. In a thin sheet of paper, similar to graphene, the energy cost of straining is much greater than that of bending. While this simple reasoning is very crude, it makes it obvious that crumpling of a membrane should lead to its softening. Furthermore, the increase of in-plane stiffness with strain seen in some devices (Fig. 2c) is also a behavior expected for a crumpled membrane since the applied stress is expected to gradually flatten the membrane and suppress crumpling (This is further confirmed via FEM, see Supplementary Fig. 2).

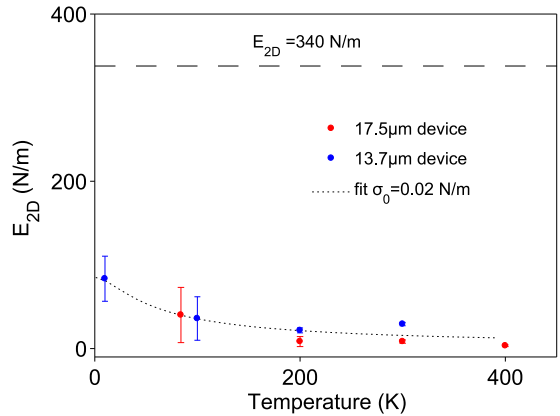

**Figure 3: Temperature-dependent stiffness of graphene.** The in-plane stiffness $E_{2D}$ measured for two circular membranes (diameters 17.5 μm and 13.7 μm) as a function of temperature. The dotted line is fit to an analytical model discussed in the Supplementary Fig. 5. The dashed line shows the stiffness of a flat graphene, $E_{2D} = 340$ N/m. The error bars are obtained by estimating the standard deviation of $E_{2D}$ measurements.

Next, we explore the relative contribution of static wrinkles and flexural phonons to the observed softening of the effective in-plane stiffness. To study the effect of flexural phonons, we examined changes of graphene's $E_{2D}$ with temperature. Since the amplitude of flexural phonons causing crumpling scales with temperature $T$ as $k_B T$ ($k_B$ is the Boltzmann constant), we would expect strong stiffening of graphene at low temperature if this were the dominant effect. We measured two different devices in the range of temperatures between 400 K and 10 K (Fig. 3). While we observed moderate stiffening of graphene from $E_{2D} \sim 20$ N/m at 300 K to $E_{2D} \sim 85$ N/m at 10 K, all of the measured devices are much softer compared to 340 N/m throughout the range of temperatures. This suggests that the contribution due to flexural phonons does not dominate the mechanics of graphene at room temperature.

To isolate the contribution due to static wrinkles, we analyzed changes in $E_{2D}$ of patterned graphene membranes. In general, there is a concentration of stress along the wrinkles in a crumpled sheet.[23] The stress can be relieved by cutting the membrane across such wrinkles. The reduction in stress, in turn, leads to a decrease in wrinkle amplitudes. In particular, for very narrow ribbons we expect fully suppressed wrinkles. This behavior is also seen in our molecular dynamics (MD) simulations (Supplementary Fig. 6). We also considered possible changes in induced strain due to chemical modification of graphene's edges and found it to be negligible for the ribbon sizes used in our study. Experimentally, our suspended graphene devices were cut using focused ion beam (FIB) lithography. The FIB beam was rastered to carve

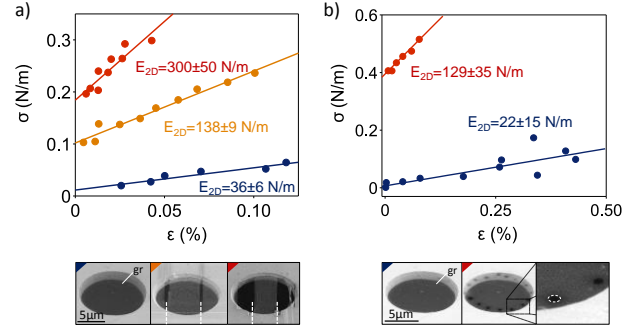

**Figure 4: Probing changes of stiffness in devices with varying geometry.** *(a)* Stress ($\sigma$) *vs.* strain ($\varepsilon$) curves for a single graphene device as its aspect ratio is changed via FIB lithography. SEM images of the device at each step of cutting are shown in bottom panels (cut directions are white dashed lines). *(b)* Stress-strain curves for another device as it is perforated near the edge of the membrane. SEM image of the device before and after perforations is shown in bottom panels. The scale bar is 5 μm.

sequentially thinner ribbons out of the same circular graphene membrane. The initial circular membrane with diameter 12.5 μm was first cut into a ribbon with width of $w = 5$ μm. The width of this ribbon was then reduced to $w = 2.7$ μm (Fig. 4a, bottom). Using SEM we confirmed that the process of cutting reorients wrinkles along the cut direction and suppresses their amplitude (Supplementary Fig. 4), this behavior is also seen in MD (Supplementary Fig. 6). We extracted the effective mechanical constants of such devices by measuring their deflection *vs.* applied electrostatic force, similar to the analysis above. For near-rectangular ribbons uniaxial stress, uniaxial strain, and in-plane stiffness were computed from known pressure $P$ and center-point deflection $h$ as: $\sigma = Pa^2/2h$, $\varepsilon = 2h^2/3a^2$ and $E_{2D} = \sigma/\varepsilon$.[24] We observed that the devices stiffen with each subsequent cut (Fig. 4a). The in-plane stiffness increased from $E_{2D} = 36$ N/m for initial circular membrane to 138 N/m for 5 μm wide ribbon, and to 300 N/m for 2.7 μm wide ribbon. The in-plane stiffness of flat graphene, 340 N/m, is within the uncertainty of the last value. We also explored an alternative approach to relieve crumpling of graphene by puncturing a series of ~100 nm diameter holes near the edge of the membrane using FIB. Similarly, we observe a significant increase in the measured in-plane stiffness after perforations (Fig. 4b). Overall, we see that once crumpling associated with static wrinkles is relieved, the stiffness of graphene increases to almost 340 N/

m. This suggests that static wrinkles have the dominant contribution to softening of the effective in-plane stiffness of circular graphene membranes.

Normally, the presence of defects lowers the mechanical stiffness of any material, including graphene.[25] However, recently it has been reported that vacancy type defects at sufficient density can lead to mechanical stiffening of graphene.[26] To confirm that the stiffening seen in Fig. 4 stems from changes in device's geometry rather than from the induction of defects in graphene that can occur during FIB cutting, we performed an additional test. To accomplish this we induced defects in suspended graphene membranes similar to the ones used elsewhere in the manuscript using irradiation with controlled dosage inside an FIB setup (see Supplementary Figure 7 for details). We then took 14 devices through several successive steps of irradiation gradually increasing the defect concentration (see Supplementary Figure 7 for details) from 0 to $\sim 5 \times 10^{13}$ cm$^{-2}$, comparable to that Ref. 26. The mechanical response of each device at each defect density was determined at room temperature using the techniques described earlier in the paper. Figure 5 summarizes our data by showing the evolution of the in-plane stiffness *vs.* defect concentration for all devices. Every device remained softer than 340 N/m in the entire range of induced defect concentrations, and the in-plane stiffness did not appear to be strongly affected by the presence of defects (apart from small changes that could be ascribed to variation in experimental conditions). This confirms that the changes in stiffness observed in Fig. 4 can only be caused by changes in device's geometry.

## Discussion

It is instructive to estimate relative contributions for flexural phonons and static wrinkling to the in-plane stiffness of our devices. The in-plane stiffness of graphene $E_{2D}$ measured in the experiment can be loosely approximated as:

$$E_{2D}^{-1} = E_{latt}^{-1} + E_{flex}^{-1} + E_{wrin}^{-1} \quad (2)$$

where $E_{latt} \sim 340$ N/m, $E_{flex}$ and $E_{wrin}$ are contributions to stiffness from three different mechanisms – stretching of carbon-carbon bonds, flexural phonons, and static wrinkles. The data in Fig. 4 suggests that suppression of the contribution due to wrinkling increases $E_{2D}$ from 36 N/m to 300 N/m. Provided that the width of graphene ribbon is much larger compared to the typical wavelength of a flexural phonons, ~10 – 1000 nm, we expect that the process of cutting does not affect the contribution due to flexural phonons. We can then use Eq. 2 to estimate $E_{wrin} < 40$ N/m and $E_{flex} > 2500$ N/m from this data. In agreement with our earlier conclusion, we see that the contribution due to wrinkles dominates $E_{2D}$. A simple estimate can clarify why the contribution due to static wrinkling is larger than that of flexural phonons. The degree of crumpling of a membrane can be quantified as $\Delta A/A = (A' - A)/A$, where $A$ is membrane's area, and $A'$ is the area of membrane's projection onto a plane parallel to it (Fig. 2d). Stretching caused by an external stress gradually flattens the membrane. When stress is large enough to suppress crumpling and flatten the membrane, the projected area is fractionally increased by $\Delta A/A$. This corresponds to a fractional increase $\varepsilon_t = \Delta A/2A$ in the linear dimensions of the membrane. We therefore conclude that when graphene is extended less than this threshold strain $\varepsilon_t$, it is mostly crumpled and should appear soft, while at strains above $\varepsilon_t$ it is mostly flat and should have in-plane stiffness close to $E_{2D} \sim 340$ N/m (in agreement with FEM, see Supplementary Fig. 2).

The degree of crumpling of graphene due to flexural phonons can be estimated using a model based on the application of the equipartition theorem to the membrane bending modes.[27] The modification of this model (Supplementary Fig. 5) taking into account renormalization of the bending rigidity at small wavevectors stemming from non-linear coupling between bending and stretching modes[28] yields: $(\Delta A/A)_{\text{flex}} \sim 0.5\%$. This estimate agrees with $\Delta A/A$ extracted from MD (Supplementary Fig. 6) and more detailed calculations.[29] The corresponding threshold strain $\varepsilon_t \sim 0.25\%$ is smaller than the average built-in strain for devices used in our experiments: $\varepsilon_0 = \sigma_0/E_{2D} \sim 0.3\%$. We therefore expect that flexural phonons are at least partially suppressed in our devices. For static wrinkling, assuming sinusoidal wrinkles with wavelength $\lambda = 50 - 100$ nm and amplitude $\delta = 1 - 2$ nm (Fig. 2d) we estimate: $(\Delta A/A)_{\text{wrin}} \sim \pi^2 \delta^2 / \lambda^2 > 0.1 - 1.6\%$. The lower bound here is likely a very conservative estimate as it neglects wrinkles with longer wavelengths. The corresponding $\varepsilon_t$ from this estimate is larger than the average built-in strain observed in experiment, and we therefore do expect softening of graphene due to static wrinkling.

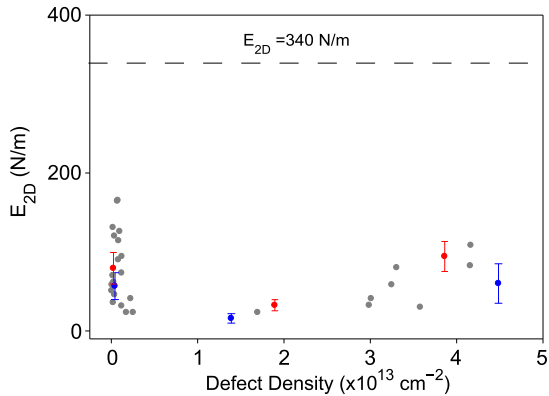

**Figure 5: The influence of defects on the mechanics of graphene membranes.** In-plane stiffness $E_{2D}$ measured for 14 different devices with varying concentration of defects. The defect density was determined via ex-situ Raman spectroscopy. For clarity we only highlighted two devices (red and blue points) whilst the remaining devices are shown as grey points. The error bars are obtained by estimating the standard deviation of $E_{2D}$ measurements.

We note that it is tempting to interpret the stiffening of graphene at low temperature seen in Fig. 3 as a signature of the temperature-dependent suppression of crumpling due to flexural phonons. Indeed, the data in Fig. 3 can be fit to an expression for in-plane stiffness obtained using the model based on Ref. 28 discussed above. This model (described in detail in the Supplementary Fig. 5) along with a realistic value of built-in stress $\sigma_0 = 0.02$ N/m fits our data well (dotted line in Fig. 3). However, since the contribution due to static wrinkling may also be temperature dependent, this agreement may be accidental.

Finally, it is interesting to compare our results with very recent relevant studies.[26,30-32] In Refs. 26, 30 and 32, the mechanics of graphene seem to be dominated by carbon-carbon bonds and flexural phonons different from samples dominated by static wrinkles considered here. In Ref. 26, the observed increase in stiffness of damaged graphene was associated with the suppression of long-range flexural phonons. In contrast, the mechanical response of our samples is dominated by static wrinkling and is not affected strongly by the presence of defects (Fig. 4 and 5). In Ref. 30, stiffening of graphene has been observed for high built-in strain ($\sim 0.6\%$). In comparison, the present work is concerned with regime of smaller strain, when crumpling of graphene is not fully suppressed. Finally, in Ref. 31 strong renormalization of bending rigidity of graphene cantilevers was observed and was attributed to both static wrinkles and flexural phonons. This is in agreement with our conclusions.

In conclusion, we have developed a non-contact technique for probing the mechanical properties of graphene (and potentially any conductive 2D material) with uniform loading and at cryogenic to room temperatures. We have confirmed that graphene is significantly softened by out-of-plane crumpling. Moreover, we developed an approach to test relative contributions of flexural phonons and static wrinkles to the in-plane stiffness of graphene, and found that the latter dominates. Our observations reinforce the idea that great care is needed when applying classical elasticity theories to atomically thick materials.[33] Crumpling (either due to flexural phonons or static wrinkling) is a salient feature of any graphene or other 2D material membrane at finite temperature.[34] The results reported here are therefore relevant for the majority of the experiments dealing with such membranes. Changes in the effective stiffness reported here should affect operation of graphene nanoelectromechanical devices including resonators[35], mass sensors[36], and switches.[37] We believe that the modification of effective elastic constants should carry over to any other thin wrinkled membrane – ranging from aged skin to solar sails.[38,39] Going forward, it would be very interesting to extend our experiments to narrow graphene ribbons at low temperatures to accurately probe the contribution of flexural phonons to graphene mechanics.

## Methods

**Fabrication of suspended graphene membranes:** Silicon nitride membranes (thickness 1 μm) were fabricated by depositing low-stress silicon rich silicon nitride on both sides of a silicon chip. An array of holes ranging between 7.5 and 30 μm were then patterned in the nitride using standard fabrication procedures. A metallic contact (Ti/Au, 10 nm/30 nm) was deposited onto the top surface of the nitride membrane. Monolayer graphene was then transferred onto holes in the nitride membranes. Different transfer procedures were used for CVD and exfoliated graphene. For CVD graphene, we use a high quality atmospheric growth and wet transfer.[40,41] For exfoliated graphene, a co-polymer stamp method is used.[42,43] Both CVD and exfoliated samples are subsequently annealed in an Ar-$H_2$ environment at 350 $^{\circ}$C. The fabrication yield for intact suspended graphene membranes varies from 55% for our smallest (7.5 μm diameter) devices to <8% for the biggest (30 μm diameter) devices. The graphene membranes remained clamped to the sample chip via van der Waals interactions forming suspended circular graphene membranes.

**Interferometric Profilometry:** A Wyko 9800 interferometric profilometer equipped with a 20x "through transmissive media" objective (NA=0.28) was used to perform optical measurements. In measurements of graphene deflection, phase shift interferometry (PSI) mode was used with HB-LED at ~530nm wavelength as illumination source. To measure graphene/gate separation, vertical scanning interferometry (VSI) mode was used with white light illumination source.

**FIB Lithography:** The FIB lithography was carried out using a Novalab 600 Dual-Beam (electron/ion) FEI. The system is aligned to the graphene with the electron beam (5 KeV, 0.4 nA) whilst cuts are made with $Ga^+$ ion beam (30 KeV, 50 pA current, approximate exposure time < 500 ms).

**Acknowledgements**
We acknowledge enlightening conversations with Paul McEuen, Isaac Storch, and Caglar Oskay as well as financial support from NSF CAREER 4-20-632-3391, Defense Threat Reduction Agency Basic Research Award # HDTRA1-15-1-0036, and the Sloan Foundation. A part of this this research was conducted at the Center for Nanophase Materials Sciences, which is a DOE Office of Science User Facility. This research used resources of the National Energy Research Scientific Computing Center, which is supported by the Office of Science of the US Department of Energy under Contract No. DE-FG02-09ER46554 as well as resources of the Oak Ridge Leadership Computing Facility at the Oak Ridge National Laboratory, which is supported by the Office of Science of the U.S. Department of Energy under Contract No. DE-AC05-00OR22725.

**Author Contributions**
Device design and actuation was devised by RJTN, HJC, NVL and KIB. Suspended graphene devices were fabricated by RJTN, HJC, NVL and IV. Profilometry data was taken by RJTN and HJC under the advisement of NVL. FIB cutting procedures were developed by RJTN and HJC under the advisement of NVL. MD modelling was performed by YSP under the supervision of STP. RJTN and VPS performed FEM modelling and AFM measurements. RJTN, HJC and KIB analyzed the data; RJTN and KIB co-wrote the manuscript with input from all authors. KIB supervised the project. All authors discussed the results.